\documentclass[sn-mathphys,Numbered]{sn-jnl}
\usepackage{graphicx}%
\usepackage{multirow}%
\usepackage{amsmath,amssymb,amsfonts}%
\usepackage{amsthm}%
\usepackage{mathrsfs}%
\usepackage[title]{appendix}%
\usepackage{xcolor}%
\usepackage{textcomp}%
\usepackage{manyfoot}%
\usepackage{booktabs}%
\usepackage{algorithm}%
\usepackage{algorithmicx}%
\usepackage{algpseudocode}%
\usepackage{listings}%

\begin{document}

\title[Quantum collapse in Everett multiverse]{Quantum collapse as undecidable proposition in an Everettian multiverse.}

\author*[1]{\fnm{Fabrizio} \sur{Tamburini}}\email{fabrizio.tamburini@gmail.com}


\affil*[1]{\orgname{Rotonium - Quantum Computing Research}, \orgaddress{\street{Le Village by CA, Pz. G. Zanellato, 23}, \city{Padova}, \postcode{I-35131}, \country{Italy}}}

\author[2,3,4]{\fnm{Ignazio} \sur{Licata}}\email{ignazio.licata3@gmail.com}

\affil*[2]{\orgname{Institute for Scientific Methodology (ISEM)}, \orgaddress{\city{Palermo}, \country{Italy}}}
\affil*[3]{\orgname{School of Advanced International Studies on Theoretical and Nonlinear Methodologies of Physics}, \city{Bari}, \postcode{I-70124}, \country{Italy}}
\affil*[4]{\orgdiv{International Institute for Applicable Mathematics and Information Sciences (IIAMIS)}, \orgname{B.M. Birla Science Centre}, \state{Adarsh Nagar}, \orgaddress{Hyderabad -- 500 463}, \country{India}}

\abstract{Our representation of the Universe is built with sequences of symbols, numbers, operators, rules and undecidable propositions defining our mathematical truths, represented either by classical, quantum and probabilistic Turing Machines containing intrinsic randomness.
Each representation is at all effects a physical subset of the Universe, a metastructure of events in space and time, which actively participate to the evolution of the Universe as we are internal observers.
The evolution is a deterministic sequence of local events, quantum measurements, originated from the local wavefunction collapse of the complementary set of the observers that generate the local events in the Universe.
With these assumptions, the Universe and its evolution are described in terms of a semantically closed structure without a global object-environment loss of decoherence as a von Neumann's universal constructor with a semantical abstract whose structure cannot be decided deterministically a-priori from an internal observer.
In a semantically closed structure the realization of a specific event writing the semantical abstract of the constructor is a problem that finds a ``which way'' for the evolution of the Universe in terms of a choice of the constructor's state in a metastructure, the many-world Everett scenario from the specific result of a quantum measurement, a classical G\"odel undecidable proposition for an internal observer, exposing the limits of our description and possible simulation of the Universe.}

\keywords{Quantum Measurement, semantic closure, Von Neumann universal constructor}

\maketitle 

\section{Introduction}

The concept ``Universe'' is mainly used to represent what defines the set of all phenomena, the physical world expressed in terms of events and coincidences of events that generate space and time and their relationships. 
The concept of event in the Universe from a classical relativistic point of view is
described in terms of a subset of the Universe for which are defined space and time properties involving those of the fundamental fields, with the meaning of ``present'' defined by its ``past'', following the Mach's principle, and generating the ``future''.
Any set of events and the possible evolution is a potential product of interactions of subsets that build up the Universe, including our mathematical truths that are metastructures made with relationships of events built with physical processes.

With these assumptions, the Universe and its evolution are described in terms of a semantically closed structure that describe interactions between its subsets, sets of events, becoming a semantically closed structure without a global object-enviroment loss of decoherence. 
A semantically closed structure describes a system that can enclose its meaning within itself. Following \cite{clark,pattee}, in a semantically closed structure of sequences of symbols, the Universe interacts with the description of itself at a given point of its symbolic sequence to replicate and evolve by copying the information encoded in its abstract description (its state and evolution at a given point of the sequence) to construct itself and its evolution with the set of rules there encoded. The sequence of symbols and rules that defines the information in the abstract description is itself encoded on that abstract description with its evolution.

In this structure each of these subsets concur as constructors or parts of the constructor for the evolution and existence of subsets up to the whole Universe by their reciprocal interaction through coincidences of events as in von Neumann's constructor \cite{voneu}. In a certain way, ideally deposed from the center of the Cosmos by the Copernican revolution, our vision of the world expressed in terms of formulation of the laws of physics and information still remains as a Ptolemaic fulcrum for the knowledge process through cause-effect relationships that relate classical macroscopic systems and the physics of quanta formulating what we call laws of physics. 

These relationships are expressed with our basic mathematical and logical language, in terms of rules, concepts, axioms and theorems in order to build a representation of nature whichever the physical Universe were finite, infinite or belongs to a Multiverse as subset including G\"odel's undecidability theorems \cite{godel} that arise when one adopts a complete mathematical description of the Universe in terms of numbers and relationship between numbers as initially proposed by Dirac \cite{dirac}. 

Many attempts have been made to relate the problem of a quantum measurement with G\"odel's undecidability, whether a proposition is true, false or undecidable comparing the limits in the formal language with the nonclassical properties of the quantum world.
From our assumptions, the measurement process is at all effects a physical process in the universe and a key for its evolution described with a sequence of symbols. If the Universe is a pure state, any measurement would affect the state of the Universe and evolve it, leaving it as pure state, mixing the states between observer and observed introducing decoherence on the observed that can be the complementary of the observer in the universe. 

The state remains a pure state if the observer is the empty set. Thus, the evolution of the universe depends on the interaction between its subsets and the problem of a measurement with a classical observer, viz., translating the evolution in terms of a formal classical language finds a ``which way'' for its evolution in a Many-Worlds Everett scenario
as a classical G\"odel undecidability proposition formulated inside the substructure Universe \cite{everett1,everett2,everett3} and depends only on the mathematical language and modeling used to describe the problem that is being faced.

Let us introduce G\"odel's Theorem with few words. There are many excellent expositions of G\"odel's undecidability theorems, and here we will limit ourselves to offer an intuitive illustration useful for introducing his projections on theoretical physics. At the beginning of the 20th century, Hilbert's program on the axiomatization of mathematical theories triggered intense research to describe at least a particular class of formal systems, those that were sufficiently powerful, through a logical syntax. These are those systems that are something more than a simple logical ``toy'', and which have a structural complexity at least equivalent to that of natural arithmetic. Practically all interesting formal theories, including physical-mathematical ones, fall into this category. Sufficiently powerful systems have a notable self-referential capacity, that is, they are able to produce propositions that concern their internal structure, on which their "fecundity" depends. This is the keystone of G\"odel's theorems in 1913 \cite{godel}, which set very precise limits to the Hilbert program as it is developed e.g. in the powerful ``Principia Mathematica'' by Russell and North Whitehead (1910 -1913) \cite{principia}, showing that formal systems ``pay'' for this great expressive capacity with logical complications of a radical nature:

\textbf{G\"odel's first theorem:}\textit{ Every sufficiently powerful, coherent and axiomatizable system is syntactically incomplete.}
\\
This result expresses that it is always possible to produce, starting from a system of axioms A, an undecidable proposition P, i.e. of which it is impossible to establish, with the tools of the system, neither the truth nor the falsity. In the theoretical context offered by the Turing Machine this is equivalent to the famous halting problem: there is no general program (algorithm) which, applied to a particular pair (program-argument), is able to establish a priori whether the relative computation the couple in question will end or not;

\textbf{G\"odel's second theorem:} \textit{Any sufficiently powerful, coherent and axiomatizable system is incapable of proving a proposition that canonically expresses the coherence of a system.}
\\
In a certain sense, the limitation of this theorem is even more drastic than the first. In fact, the theorem states that using the syntactic-formal tools of <L, A, R> it is impossible to demonstrate the logical solidity of the system itself, and in particular to predict the production of a contradictory development. The hope placed by Hilbert in the axiomatic method as an instrument for the security of a logical foundation of mathematical knowledge was thus undermined at its roots.
Outside of formal languages it is possible to understand that axiomatization procedures require a typically tree algorithmic compression for each mathematical theory, which is certainly possible within the semantic world of each specific model. However, when the emphasis is placed on purely syntactic aspects, the specter of undecidability hovers over the production of new propositions because, contrary to what many people think, mathematicians do not only manipulate symbols but also meanings, building connections between different models with perfectly legitimate procedures, the first order propositional calculus is coherent, complete and decidable and generally not very risky. Meaning that the calculus of predicates, which is obtained from the propositional calculus with the introduction of quantifiers, is undecidable, but in general the semantic constraints on the discourse they protect from logical flaws. 
In other words it is legitimate to think that if undecidability comes into play even with ``elementary'' systems such as the axiomatizations of arithmetic, the situation can only become critical when developments that imply different areas are considered superimposed. A now classic example is Fermat's theorem, cited by G\"odel in a 1928 conference as a possible undecidable proposition \cite{goldstein}. The theorem was then proved by A. Wiles in the 1990s using very different branches of mathematics and completely unknown in G\"odel's time \cite{singh}. A recent example is the proposed solution of the Riemann conjecture through the equation with infinite Majorana components (Majorana tower): also in this case the construction procedures of the demonstration went far beyond the possible care in the construction of a axiomatic system; the two structures in fact belong to very different semantic fields \cite{tamlic}.

\section{Computing in a Universe}

Any formal language can be translated in Turing Machines. G\"odel's undecidability is translated in terms of the problem of computation, the Turing Machine halting problem through the Church-Turing lemma. When related to a physical system takes on radically different aspects in the axiomatization of physical theories, where the axiomatic has always been little less than an attempt at synthesis, mixing theoretical elements and empirical assumptions \cite{sneed}. The point is that physics, however formalized, is never a syntactic system because every physical theory has an intrinsic semantics defined with operational procedures. Nonetheless, it is interesting to ask whether physical theories are undecidable and what foundational questions a ``theory of everything'' can pose.

\subsection{Undecidability and uncomputability in theoretical physics}

As is known, G\"odel's Theorems constitute a fundamental stage in the relationship between logic and mathematics. Shattering Hilbert's formalist dream, the undecidability results contributed to the contemporary conception of mathematics as an open, non-zippable system \cite{chaitin}. 
Considering theoretical physics as a formal construction \cite{lic1} , it is interesting to investigate the possibility of finding undecidable propositions here too. In this sense there are some general results that depend on the observer being immersed in the system he observes like discussed in Ref. \cite{breuer,uri}, and others more specifically addressed to the questions posed by quantum physics and cosmology, where the unpredictability of a event is rooted in the nature of things. Now classic examples are the collapse of the wave function (which is a bit like the fifth postulate of QM) and the cosmological configurations of the universe described by the Wheeler de Witt equation \cite{wdw}. In both cases there is no calculation, and in particular no algorithmic procedure, capable of connecting the range of possibilities to the event or universe that we actually observe. At the quantum level, therefore, the computational pact established by Church Turing's thesis seems to collapse. One way to describe this situation is that Shannon-Turing information, of a local nature, cannot trace the evolution in detail of a quantum system as happens for a classical system because there is ``hidden information'' associated with the described entanglement, for example, through a quantum potential or Feynman integrals. The problem becomes even more radical at the level of quantum gravity because in GR the causal structure is not fixed, which added to the quantum uncertainty defines an indefinite causal structure \cite{geroch,lic2,garcia,hardy}.

Of particular interest for our purposes is to underline that the fundamental reason for the indeterminacy in quantum cosmology is of a mathematical nature and derives directly from the G\"odel limits, it is the non-classification theorem of four-dimensional varieties: there is no algorithm that can classify all compact four-dimensional manifolds not limited, nor even capable of distinguishing between two of them \cite{freedman}. This is, as is evident, an issue at the heart of the so-called "peaceful coexistence" between GR and QM and which makes it difficult to define a cosmological wave function. However, there are alternatives. For example, it is possible to choose for physical reasons a selection criterion that selects a geometry from the totality of the manifolds as a cosmological boundary condition. This, for example, is the path chosen by Hartle-Hawking and other physicists which assigns a special role to de Sitter's geometry recently confirmed by observations on the acceleration of the universe \cite{hartlehawking,feleppa}. A more extreme solution could lie in the advent of cellular automata universe models on the Planck scale, and therefore the extreme conceptual complexity of the varieties is not necessary if not coarse grain, and is replaced by a discrete algorithmic complexity. This path was taken by 't Hooft in an attempt to unite the interpretation of QM with particle physics \cite{tofft,tamlic2}. 't Hooft's interpretation is often described as an attempt to reintroduce locality to the bottom of the
QM, but this is not entirely accurate. In fact, the periodic orbits inside the cells which replace in a very precise sense the harmonic oscillators of QM cannot be observed directly (fast hidden variables), and the equivalence classes support the effects not local up to the Planck scale. It is therefore, to all intents and purposes, an emerging version of QM. Even in this version, however, the measurement event seems afflicted by the unpredictable characteristics of collapse. Indeed, if we do not limit the idea of collapse to the traditional observer-observed binomial and connect it to the more general objective concept of interaction (a measurement is an interaction) we obtain the image of a universe that is ``actualized'' through interaction events unpredictable starting from the fundamental laws of physics giving rise to emerging properties and metastructures and complexity.

One of the central issues in theoretical physics is the study of complex behaviors within systems. The notion of complexity is not singular; it spans across disciplines, justifying the multitude of possible approaches based on the peculiarities of the system under consideration. However, there is a deeper epistemological reason for this diverse landscape of the ``archipelago of complexity'': it's the pivotal role of the observer in detecting situations of complexity-instances where the collective behaviors of a system lead to structural modifications and hierarchical orderings \cite{lic2}. This consideration leads directly to the crux of the issue of emergence in physics.

In general, intrinsic emergence is when we see a discrepancy between the formal model of a system and the observed behaviors. In other words, the recognition of emergence expresses the necessity, or at least the utility, of creating a new model capable of encompassing the new observational ranges. This raises the problem of the relationship between different levels of description, leading to two possible situations. 

The first is known as phenomenological emergence, which concerns the semantic intervention of the observer regarding the new behaviors of the system. It aims to create a model whose characteristics - selection of state variables and dynamic descriptions - are aimed at a more convenient description of the observed processes. In this case, it is always possible, at least in principle, to connect the two models through appropriate ``bridge laws'', whose task is to link the two descriptive levels via a finite amount of syntactic information.

The second one is the radical emergence, which involves a completely new and different description that cannot be linked to the original model. Here, a breakdown of the causal chain is usually observed and can be described with appropriate symmetries and irreducible forms of unpredictability. In this case, the connection between the theoretical corpus and the new model may require a different type of theory semantics, such as a new interpretation and a new arrangement of basic propositions and their relationships like in statistical physics \cite{landaustat}.

These two distinctions should be considered purely illustrative, as more varied and subtle intermediate cases can indeed arise. As an example of phenomenological emergence, consider the relationship between Newtonian dynamics and the concept of entropy (via Standish). Classical dynamics laws are time-reversal, whereas entropy defines a 'arrow of time.' To bridge the two levels, a new model based on Maxwell-Boltzmann statistics and refined probabilistic assumptions, in turn centered on space-time symmetries, is necessary (due to the isotropy and homogeneity of space-time, there are no privileged points, directions, or instants in a process of energy level de-correlation). This allows for establishing a 'conceptual bridge' between particle description and entropy, and thus between the microscopic and macroscopic analysis of the system. However, this connection does not cover all aspects of the problem and cannot be seen in any way as a 'reduction.' In fact, in some cases, even within the closed formulation of classical physics, entropy may locally decrease, and no one would ever think of describing a perfect gas in molecular terms!

Another example concerns EPR-Bell correlations and the role of non-locality in Quantum Mechanics. In the Copenhagen interpretation, non-local correlations are observed but are not part of the theory's facts. In Bohm's interpretation, the introduction of quantum potential allows incorporating non-locality within the theory. It's worth noting that historically, the EPR issue originated as an 'ideal experiment' between Einstein and Bohr on the 'elements of physical reality' of QM. Only later, with Bohm's analysis and Bell's inequality on the limits of local theories with hidden variables, was it possible to transform the issue into experimental matter. Neither Einstein nor Bohr actually expected to observe 'spooky actions at a distance.' Importantly, the expression of non-locality in Bohm's theory does not require additional formal hypotheses beyond the standard framework provided by the Schrödinger equations. However, while this new interpretative perspective provides a different understanding of the theory, it also raises issues regarding what has been termed the 'peaceful coexistence' between special relativity and QM.

In both briefly examined (cited) cases, we can see how phenomenological and radical aspects of emergence are deeply intertwined in the dynamics of the development of physical theories. Moreover, it underscores the fundamental role of the observer in modeling and interpretive choices. It's essential to note that the relationship between observer and observed is not a bipolar relationship and, to prevent epistemological impoverishment, it cannot be resolved in a single direction. Instead, it should be considered an adaptive process where the system's internal logic meets our ways of acquiring information about it to construct theories and interpretations capable of providing a description of the system.

In this work, we will examine the challenges posed by an emergent theory understood as a general theory of the relationship between the observing and observed systems. We will review various approaches to emergence and test them on models of evolution of classical and quantum systems. Finally, we will develop some considerations on the logical limits of theories and the role of computability in describing the physical world

\subsection{Computing the Universe as a Whole}

The broader definition of the universe proposed is due to the medieval philosopher Iohannes Scotus Eriugena (810-877) who understood it as everything that is created and that is not created. For a modern mentality, the reference to what is not created, or cannot be created, is interesting in relation to the importance given to constraints. Today we could include in the definition the probability of an event imposed by Quantum Mechanics, and replace the theological accents with the big bang and the conditions that define space-time and physical laws. All this must be distinguished from the observable universe, whose boundaries are those we know from cosmology understood as the history of matter, and is different from the set of possibilities contemplated by theoretical physics. The push towards unification pushes the archipelago of physical theories towards a greater number of connections around some central islands (relativity, Quantum Mechanics) and some mathematical keys (gauge theories). These connections imply very strict requirements on the constraints of possibilities, to the point of suggesting the idea of a new approach to physics based on these \cite{costruttore1}.
The question we ask ourselves is whether a theory of everything can be considered as a Gödel system and which aspects of the universe would remain undecidable or incomputable. It should be underlined that in the current state of knowledge, non-computability should not be understood in terms of algorithmic compression (Church-Turing thesis) because varieties constitute a central part of much of the knowledge of the physical world; furthermore, the very structure of quantum physics poses problems for the universality of a quantum MdT \cite{lic4}.
An important ingredient of the universe is, in addition to chaos and randomness (Kolmogorov-type), the presence of organized complexity, which favors the development of structures with logical depth \cite{bennett}. In the current state of affairs it is very difficult to say whether this aspect of the universe derives from physical laws or rather from special boundary conditions, as seems more likely \cite{barrowtipler,susskind}.
We have arrived at the crucial point regarding the question of the physical world as a Godelian system, understood in a broad sense. On the one hand we can resolve the issue of the incomputability of the cosmological boundary conditions (WdW equation) by choosing a specific geometry, as in the Hartle-Hawking case; in this case the collapse of the wave function remains an unpredictable event with characteristics of randomness, an undecidable event on the basis of fundamental laws. As is known, the enigmatic aspects of the collapse dissolve in Everett's many-worlds interpretation, which has merged today with the cosmology of chaotic inflation \cite{dewitt,vilenkin,tipler}. This new powerful cosmological interpretation of QM seems capable of solving both the two undecidability problems, that of the choice of boundary conditions for the universe and the collapse of the wave function, suggesting that the multiverse is a logically closed and consistent system fusing physical laws with boundary conditions \cite{aguirre}

The Church-Turing Deutsch principle (CTD), formulated by Deutsch \cite{deutsch1,deutsch2,timpson}, states that a universal computing device can simulate any physical process; \textit{``every finitely realizible physical system can be perfectly simulated by a universal model computing machine operating by finite means''}. Any Turing Machine can in principle be built to describe any physical phenomenon. 
Turing Machines, including quantum and classical computers are at all effects also physical systems and anything they can do is dictated by the laws of physics, including our language and the building of mathematical truths.
Thus, the physical limits of computation are determined by the fundamental constants of Nature such as the speed of light $c$, the quantum of action $h$ and the gravitational constant $G$, with well-defined quantitative bounds \cite{lloyd}.

Paradoxically, following the laws of computation, ideally, the Universe could be simulated by a quantum computer or by a suitable Turing Machine with the consequence that one can deduce the possibility we can be the result of a quantum computer simulation and live inside it.

Differently from the mathematical languages in which can be defined true and false and undecidable propositions, these concepts take on radically different aspects in the axiomatization of physical theories, where the axiomatic approach has always been little less than an attempt at synthesis, mixing theoretical elements and empirical assumptions \cite{sneed}. The point is that physics, however formalized, is never a syntactic system because every physical theory has an intrinsic semantics defined with operational procedures. Nonetheless, it is interesting to ask whether physical theories are undecidable and what foundational questions a ``theory of everything'' can pose.

One of the main points is the problem of the Quantum measurement, which is at the heart of quantum information processing and is one of the criteria for quantum computation. 

These properties include also metaproducts or emerging structures from sets of physical systems with emerging laws different from the basic laws of Quantum Mechanics. An example are Classical systems that deal with the concept of real numbers, which cannot be simulated by a Turing Machine, as TM can only represent computable reals as the product of a finite calculation. 

If the Universe is finite, contained within a given finite region of space like in a sphere of radius $R$, it contains a finite amount of information and energy $E$ and thus of entropy. This is given by the Bekenstein bound, an upper limit on the thermodynamic entropy 
\begin{equation}
S \leq \frac{2\pi k R E}{\hbar c \ln 2},
\label{beke}
\end{equation}
where $k$ is the Boltzmann's constant, $\hbar$ is the reduced Planck constant, and $c$ the speed of light. 

The entropy can be described also in terms of the Shannon entropy 
\begin{equation}
H= \frac S{k \ln 2}. 
\label{entropy}
\end{equation}
In other terms, this quantity gives the maximal amount of information required to fully describe any given physical system down to the quantum level.

The information of a physical system, or the information necessary to perfectly describe that system, must be finite if the region of space and the energy are finite as expressed in Eq. \ref{beke} and \ref{entropy}.

In computer science, this implies that there is a maximal information-processing rate, the Bremermann's limit \cite{brem1,brem2,jordan,lloyd2} for a physical system that has a finite size and energy, and that a Turing Machine with finite physical dimensions and unbounded memory is not physically possible. Unless assuming that the mathematical truths are emerging metastructures, viz., structures that do not directly depend from the initial physical laws, the actual representation of integer and real numbers would not be possible in a finite Universe, including the representation of infinities, unless assuming the existence of local continuous variables that reflect the classical concepts of space and time or the quantum mechanical continuous variables. Being continuous, in the mathematical language one needs to build an axiomatic definition that includes Dedekind's cuts with elements that has to be Dedekind-complete \cite{dedekind} or with Tarski's axiomatization \cite{tarski}, that does not show a direct connection with what we call the basic physical principles.

Following the Constructor theory, information is expressed in terms of which transformations of physical systems are possible and which are impossible using the language of ergodic theory \cite{constructor}. 
An input substrate is processed by the Constructor giving an output substrate. Each event or measurement process can be expressed in terms of constructors and substrates up to a universal constructor for which input and output substrates represent the evolution of the Universe, namely, a universal constructor. In this way the Universe can be represented in the Constructor theory framework. Being a product of the mathematical language, a Constructor to represent the whole Universe must have the same set of information to build and build the evolution of the Universe.
Information is something whose nature and properties are determined by the laws of physics alone. Information is also of the essence in the preparation of a state and measurement in physics: input and output of state preparation and measurements is represented by a set of information quantities, being information a physical process.

Modern physics has recently developed some high level phenomenology models
setting a sort of end-game theory model, which needs no a priori notions, to obtain
a way of describing the system Universe as a whole, like in the Wheeler - De Witt wavefunction of the Universe, in the view of describing the Universe in terms of a self consistent semantically closed object.

Axiom-based models generate object-based logic and meta
structures with their composition meta-rules, and are based upon symbols acting as 
fictitious objects obeying to some meta-rules, requiring an infinite hierarchical
regress to higher level modeling, capable to include building from any emergent 
phenomenon. This is an aspect derived from G\"odel theorems for a formal logical
system that can be, under certain hypoteses, classically translated into a Turing
Machine by the classical Church-Turing thesis: each computable function can be 
computed by a universal Turing Machine.
While generalizing this universality to quantum computation, there should be
a universal quantum Turing Machine performing any desired unitary
transformation on an arbitrary number of qubits, including a program as
another part of the input state; or the program effecting a unitary
transformation is independent of the state of qubits to be computed. It is
shown, however, due to entanglement, neither of these two situations exists
in Deutsch's quantum Turing Machine \cite{dtm}. In this case an input state is
unitarily evolved to an output one. Such an algorithm, written in classical
language, consists mainly of two parts:

{\em 1.} how to embed the problem in the input state and the result in the
output state,

{\em 2.} how to realize the desired unitary transformation in terms of
various quantum gates and wires, i.e. how to construct a quantum
computational network. As shown in \cite{yushi} the Church­Turing thesis,
cannot be, as it is, generalized to quantum computation, i.e. an arbitrary
unitary transformation can be realized by a network composed of repeated
application of local operations of gates and the algorithm for composing the
network is classical. We have two types of universality in its quantum
generalization:

{\em Type­-I} universality refers to the ability of performing any desired
unitary transformation on an arbitrary number of qubits, by including a
program as another part of the input state, similar to the classical one.

{\em Type-­II} universality means that a same program can be used for
different input data.

Linearity and unitarity of quantum evolution conflict these two types of
universality in Deutsch's QTM, and the two types of universality in quantum
computation as possible generalizations of the notion of universality in
classical computation as stated in the ChurchTuring thesis doesn't exist,
because with dynamics fixed, linearity and unitarity of quantum evolution
makes it impossible to synchronize different quantum paths of any
possibility. This difficulty originates in entanglement. For a specific
quantum computation, however, there is not such a difficulty by definition;
Church-Turing thesis is interpreted as a physical principle, related to the
problem of quantum measurement. 

To model without a priori notions and infinite nesting into metastructures,
start up axioms must be given via universality property by self-organised criticality.
This describes the property of many systems to self organise in such a way that 
the system itself moves towards states characterised by a fractal-like  description,
with no fundamental local scale.
It had been shown in \cite{cah}, generalizing G\"odel and Chaitin results in mathematics,
 that self referential and self contained systems as the universe must involve intrinsic 
non local randomness, namely self-referential noise as a realisation-independent 
characterisation of self-referencing. 
Recent developements on nonlinear quantum field theory, also according to prescriptions 
of Nelson's Quantum Mechanics and stochastic quantisation, had recently shown that also 
a simple $SO(10)$ model could generate stochastic behaviour during its evolution 
\cite{brufab} giving universality to chaotic inflationary scenario described via an 
interacting many field model.
This ''noisy'' property is a general feature of quantum scenario: Wiener processes and 
Fractional Brownian Motions characterize each model evolution with a specific fractal 
behaviour depending on its global particle statistics in a generalized Haldane scenario \cite{haldane,lic5}.
Chaitin stated that if the system is sufficiently complex, the self-referential capability 
of arithmetic, results in randomness and unpredictability in a thermodynamical sense of view. 
Local randomness arises also from quantum measurement processes, which is shown to be an 
undecidable proposition inside the structure universe, and should require a metastructure 
as Everett scenario in which must be defined as a universe-choice process following the model of Nelson's Quantum Mechanics \cite{nelson,nelson1}. 
Using a classical formulation brung up from nonlinear dynamics, in a finite time computation 
model (i.e. locally), quantum and classical nonolinear system are undistinguishable, and the 
choice corresponds to an arbitrary stop of a classical Turing Machine.

The Heisenberg's uncertainty principle in this scenario could act as geometric
self-similarity operator, and Mach principle is intended as measurement
operator, essentially given by the nonlinearity property of the fields
acting as a semiclassical object interacting with a quantum one (as it
happens during reheating when the inflaton, described semiclassically
because of its is interacting with other quantum scalar fields). 
So, the self reproducing universe brings its existence by its self interaction.
Showing that the problem of quantum measurement is closely related to
G\"odel thesis, this could imply as {\it necessary} condition the existence
of a  metastructure, or of a self referential structure given by self-reproducing and 
organized hyerarchical interacting baby-universe scenario geometrically described, as a scaling characteristic of its fractal property by Heisenberg uncertainty principle: Noether's theorem.
The (self)generation processes of the universe are such as to guarantee a quantum tessellation (with Generalized Uncertainty Principle) and then an emergence of what we call isotropic and homogeneous spacetime. Noether! And there is no clear difference between the universe and the multiverse, especially in QM ("inside" and "outside" depends on the degree of logical openness of the model, which cannot be infinite, i.e., what is valid for mathematical metastructures it does not necessarily apply to physical events, which also gave rise to formalisms. Among other things, the development of quantum gravity will lead us better to define these aspects suspended between continuous and discrete).

An example can be found in the inflationary scenario Linde and Vilenkin \cite{linde1,vilenkin1,vilenkin2,vilenkin3} discussed the possibility, under certain hypotheses, about the existence of a self-reproducing inflationary universe described as a self-reproducing fractal structure. In this context, this scenario could be succesfully seen as belonging to the class of what we call ''meta Everett Universe'', without needing a specific initial choice of parameters, interpreted as an emerging self-semantically closed structure which reproduces itself during its evolution.

Some properties of the language able to discuss this sort of semantically
closed self-objects are classically well described by second order cybernetics 
concepts; as an example, von Foester \cite{vonfoe,rocha} postulated the existence of solutions for an indefinite classical recursive equation derived from Piaget recursive structure of implications which describes an observer's account of a causal interaction between observer and observed without any starting point and with event ordering property, $Obs_t=Coord(Obs_{t-1})$, i.e. a chain of implications into the self-referential structure that defines ordering processes inside the structure, as stated by von Neumann. 
The solutions to this symbolic equation represent a stability structure in terms
of discrete eigenvalues $O_i$ into the chain of infinite implications and act into 
the formal logic structure as a group of axioms for a metamodel, i.e. a model of 
modeling the reality, in which all fundamental properties can emerge from a self-organizing process of the structure universe itself, by means of a process named eigenbehaviour. 
A self organizing system like this, is autonomous \textit{iff} (if and only if) all structural processes that define and sustain all its dynamics are internally produced and reproduced, with an organizational closure, as seen above. Changing its classification ability, is shown that the system must change its own structure. 
Following this prescription is possible to define the problem of quantum measurement as a process, i.e. an object inside and belonging to the universe itself, the result of an ensemble of physical processes that concur to create a global one which changes its own structure defining a cosmic time, keeping the energy constant, according to thermodynamics and the decoherence problem of a pure state (in fact, if we make a distinction between observer and observed, the transformation of a pure state as the universe, into a mixed one by a measurement process, is described by the change of Hamiltonian mean value, in violation of the energy conservation principle). 

Thus, each physical process can act as a measurement process to its complementary into the universe, giving rise to a surging up of ''infinite'' state superposition concurring to describe the evolution of the global state universe, as the result of a collapse of all those possible states of existence. 
The problem with infinities is so linked with Cantor's theorem that denies 
bijections between $\mathbb{N}$ and $\mathbb{R}$, or the classical Turing Machine stop for non computable functions. 

In this way, roughly speaking, the collapse and the quantum measurement should be removed from the aura of ``mystery'' once and for all; in the end, we simply cannot count all the interactions in the universe (which we should know since Feynman paths and Bohm's potential \cite{lic5}. With Everett (and in this part our meaning is strengthened and clarified), Rovelli and Von Foerster claims we build our ``eternal brilliant garland'', also removing the na\"ive residue of decoherence. Without collapse there would be no particles!
What remains, up to the conclusions, brings to fruition the link between physics and computation. In practice, we have removed Everett and decoherence from the banal readings of the ``fundamentals'' bringing them back to the center of concrete physics. A rough synthesis could look like this: Every nucleation from the quantum vacuum, i.e. production of the universe, is so rich in complexity (viz., metastructures) that for every observer-language there will always be physically undecidable propositions.

In principle, this can be described by a Heraclitean Process System (HPS) with self-organizing critically characteristics: randomness, nonlinearity, nonlocality and 
iterative structure to give rise to a fractal like structure. 
The linear iterative map is given by the results of a local quantum mechanical measurement, generating self referential noise. It is a manifestations of HPS characteristics via objectification processes, as nonlocality configurations induced by macroscopic objects, or nonlinearity itself, as Mach principle states.
Nonlinearity behaves as a macroscopic semiclassical apparatus, giving discrete jumps. 
Some approaches to this statement had been done by defining a quantum measurement as a sequence of binary quantum jumps caused by a macroscopic apparatus \cite{mash,mash1}, avoiding the creation of a perpetuum mobile of the third kind \cite{proof5b}; in this way an additional structure of the spacetime is added, in terms of an hypersurface with a constant value of what is interpreted as cosmic time. In the Heisenberg picture the state vector changes at each quantum jump, which corresponds to a spacelike hypersurface. 
Each quantum jump gives a disjoint hypersurface, causewise ordered as in von Foerster equation: 
$S_2 > S_1$ or $ S_1 > S_2$ . 
A cosmic macroscopic time for the instantaneous wavefunction collapse is defined, and time 
gets in this scenario a precise meaning as any other quantum result. 
When applying the measure to the state universe, we must define an operator capable to 
describe the collapse of a state inside Everett scenario. The description if a quantum 
measurement operation has occurred or not, is given by Rovelli operator $M$, \cite{rovelli}, 
which has the crude meaning of ''{\it It has happened or not}'' and logical
eigenvalues $"1"$ or $"0"$. 

To give an example, when applied to a physical system whose observable can be usually expressed in terms of state projectors as: 
\begin{equation}
A = \sum {a_i P_i} ,
\label{observable}
\end{equation}
where $P$ is the general projection operator on a given basis. 
In our case we consider the Universe in a split between an observer, a part of the Universe and the complementary set in the Universe of the observer from the point of view of the internal observer.
Let us consider the system vector under measurement $\varphi $ and $\xi $, the vector-state apparatus, then $\psi $ represents the composed system between them after an interaction.
Before the measurement process we have $\psi _{bi}=\varphi \otimes \xi $, then the state becomes 
$\psi _{ai}=\varphi _i\otimes \xi _i$ with probability 
$\omega _i=(\varphi ,P_i\varphi )$ and $\varphi _i=\frac{P_i\varphi }{\left\|P_i\varphi \right\| }$. 

From the definition in Eq. \ref{observable}, their statistical operators are defined in the following way, 
\begin{eqnarray}
&&\rho_b^{comp}=P_{\psi _b}^{comp} \nonumber
\\
&&\rho _a^{comp}=\sum\omega_iP_{\psi_{a_i}}^{comp}
\label{composed}
\end{eqnarray}
that correspond to the vector $\psi $ of the composed system, in agreement with the energy conservation principle and with laws of thermodynamics. 
By the concept of decoherence, the pure state Universe is seen as the composed system of Eq. \ref{composed} and is equivalent to the one given after the measurement in terms of state projectors:

\begin{eqnarray}
&&\rho _a^{\symbol{126}comp}=P_{\psi _a}^{comp} \nonumber
\\
&&\psi _a=\sum (P_i\varphi)\otimes \xi _i\Rightarrow \rho _a^{\symbol{126}comp}= \rho _a^{comp}
\label{composed2}
\end{eqnarray}
with the decoherence condition 
\begin{equation}
(\psi _{aj},O\psi _{ai})=(\varphi _i\otimes \xi _i,O\varphi _i\otimes \xi _i)=\delta _{ji}. 
\label{deco}
\end{equation}
To avoid the decoherence condition of Eq. \ref{deco} and obtain the pure state universe, following Eq. \ref{composed2}, one has to make undistinguishable the pure state from the mixed one embedding all into a
structure that must evolve after each process has occurred, changing its
structure in terms of interactions between its subsets. 

We can state that each event concurs to the evolution of all universe and all universe generates
each event, as Mach principle states. This defines the Universe in terms of a von Neumann's universal constructor \cite{voneu}. 
To give a simple example, if the measure of $A$ is described with two discrete eigenvalues only, e.g. $a_1$ and $a_2$ with eigenstates $|a_1\rangle $ and $|a_2\rangle $ and interaction System-device Hamiltonian $H_I$. 

Following the prescriptions of Quantum Mechanics , we prepare the state of the device $|init\rangle $ such that in a finite time the interaction will evolve into $|a_1\rangle \otimes |init\rangle $ into $|a_1\rangle \otimes |\xi a_1\rangle $ and $|a_2\rangle \otimes |init\rangle $ into $|a_2\rangle \otimes |\xi
a_2\rangle $, i.e. $\psi _a=\omega _1\varphi _1\otimes \xi _1+\omega
_2\varphi _2\otimes \xi _2$.

Then, $\psi (0)=(\omega _1|a_1\rangle +\omega _2|a_2\rangle )\otimes |init\rangle
\longrightarrow \psi (T)=\omega _1\varphi _1\otimes \xi _1+\omega _2\varphi
_2\otimes \xi _2$ is a pure state that is replaced with one of the two
substates after the wavefunction collapse $\psi _1=\varphi _1\otimes \xi _1$
or $\psi _2=\varphi _2\otimes \xi _2$. Computing the probabilities for the
collapse, we have a correspondent hypersurface for any given state $\psi (t)$
of the combined state observer-observed $(\varphi \otimes \xi )$ system by
means of Rovelli correspondent operator $M\equiv |\psi _1\rangle \langle
\psi _1|+|\psi _2\rangle \langle \psi _2|$ acting in this way:

State\ ''happened'' with eigenvalue $1$

$M(\varphi _1\otimes \xi _1)=\varphi _1\otimes \xi _1$and $M(\varphi
_2\otimes \xi _2)=\varphi _2\otimes \xi _2$

State\ ''not\ happened'' with eigenvalue $0$

$M(\varphi _1\otimes \xi _2)$ and $M(\varphi _2\otimes \xi _1)$

In this way, time and measure and event take sense into the Everett's Many-Worlds model, in which every instant of time corresponds to a special case of a universe, and a measure with its result is described as a causal process with an ordering operator capable to define a cosmic time. Only in this meta-universe the problem of decoherence can be translated as a problem of universe choice: phase coherence is lost
by physical decoherence, into the environment, making pure and mixed state indistinguishable  and changing the structure of the global state in terms of mutual relationships between its substates, that means time evolution. The complementary of an observer can in fact be seen as the environment surrounding a quantum system. In this way the observer can monitor the observables, or part of them, of the system. The effect of the observer is to induce decoherence continuously in the eigenstates of these observables and can assume the behavior similar to that of classical or pseudo-classical states, \cite{zur,zur1,zur2} as in the convergence of quantum Turing Machine into the final state of computation.

For this reason the Everett scenario seems to become a sort of necessary environment for the evolution of the pure state that represents the object universe when described in terms of classical (or pseudo-classical) formal propositions with the possibility of building a G\"odel symbolic construction (or a Turing Machine) to describe the evolution of the universe and the interaction between subsets up to the interaction with ``classical'' (or pseudo-classical) observers that leads to the problem of a quantum measurement to be related to an undecidable proposition inside the linguistic representation of the universe itself.
The implication is to have as mandatory the existence of a metastructure as a ``linguistic'' meta-universe. 
Being this representation of the meta-universe a class of universes in the evolution of our Universe and at the same time built with physical events in the Universe, the class should be a sub set of the Universe by definition or coincide with the Universe, recalling Russell's paradox.
The mandatory requirement of the existence of a universe of universes, shows that our classical  language is inadequate for describing the state universe as a whole self-object unless employing a self bootstrapped structure, in which the laws of physics self sustain one another through their mutual consistency \cite{baumann}. 

\section{Setting a language toy-model}

We now show with a simple toy-model that the description of the Universe in terms of Turing Machines and constructors with our mathematical tools and truths, this unavoidably leads to relate the events in the Universe and its evolution with the undecidable prepositions in our language following point-by-point the logical structure set by of G\"odel to define the undecidable propositions in a logical structure.

To proceed in this paradox, one of the further steps to do is the generalization of the Turing Machine to describe continuous spectra, translating the construction of a hypersurface quantum jump in usual terms of either probabilistic or quantum computation. In this case is possible to build a integer-number codification of a 
obtaining ordered sequences of numbers and numerals, which correspond to physical 
quantities and laws, seen as numerical relations defined into the structure universe itself.
An example is the adimensional Dirac \cite{dirac} construction of physics, where numbers express physical quantities and relationships between numbers represent laws:

1) Event, causally measured physical quantity as a number

2) Laws, relationships between quantities, class $\rightarrow $ numeral

3) measure, coincidence or relationship between events, class $\rightarrow $ 
numeral

What is observed, or physically defined, is the result of an indefinite succession of cognitive (cause-effect) interactions that describe all the possible sub splitting of the pure state universe in a superposition of
mutually interacting subsets. To obtain a formal-logic description of the universe, the self evolution of an object with ordered time as internal state must be described, and must be used proper elements of the system to
describe relationships between other elements belonging to the system inside the system itself, as a self-referential structure. According to G\"odel's undecidability theorem \cite{godel}, it is impossible to show formal coherence of the structure inside the structure itself. 
The a-priori choice of the result of a quantum measurement, expressed through the classical logical language, as discussed before, becomes an undecidable proposition inside the universe itself, and takes sense only
into Everett scenario. 
The problem of quantum measurement, as defined, becomes an undecidable proposition inside the structure Universe itself; for this purpose some tools, variables and classes are defined, as in G\"odel \cite{goedel}. This assures that assertions are valid also for transfinite systems: primitive concept of ''{\it follows}''  by operator $f$ ; $-,\symbol{126}$ are ''Not'', $\vee $ Or, $\equiv $ equal by definition $(x)\equiv \forall $ for each; $\supset \subset $ iff; $(x),(Ex),(\varepsilon x)$ with a border for the variable x, are used in definitions and propositions to express that the concept defined there are recursive. 

Following  any logical system as a physical model is, corresponds to a meta-(...)-metastructure of integer
numbers $Z$ going from a logical formulation of higher order, justifying certain undecidable propositions, and shortening in a pretty large way an ideally infinite number of other proofs, but they do contain undecidable
propositions leading to the Russell's paradox.

If one builds a sequence for a Turing Machine to describe the phenomena in the Universe and the problem of measurement then formally obtains the same representation of a Turing Machine and the G\"odel undecidable propositions:

I - logical symbols in term of constants \{$\symbol{126},\vee ,\forall ,\ 0,\
f\ ,(,\ ),,...$\};

II type-1 variables (numbers as quantities, zero included) ; type-2
variables (classes of quantities); type-3 variables (classes of classes...)
; type-4(...) variables (classes of...)... and so on.

{\em def 1:} $Flg(k)$ is the set of consequences of $k$, the smallest set
containing all $k$-formulas, axioms, and is closed respect to the
relationship $f$ of immediate consequence (e.g. in terms of hypersurfaces)
(G\"odel functions $43,TheoremV$) \cite{goedel}

{\em def 2:} $a$ is a sequence of numbers, i.e. a formula.

{\em def 3:} $\nu $ is the free variable of $a$

{\em def 4:} $Z(n)$ is the numeral of $n$ respect to which the proposition
is done

{\em def 5:} $Su\ a(_{Z(n)}^\nu )\equiv Sb(a_{Z(n)}^\nu )$ substitute in $a$
of $Z(n)$ to the $\nu $-term of $a$ (G\"odel functions $27,30$)

{\em def 6} $\nu Gena$ is the generalization of $\nu $ respect to the
variable $a$ if the last one is a variable (G\"odel function $15$)

{\em def 7:} $Not(x)$ not-x (G\"odel function $13$)

{\em def 8:} $S$ is $\omega $-non-contradictory ($\omega $-coherent) iff for
no property $F$ of natural numbers is not possible to demonstrate if a
formula is true or false, i.e. $(Ex)\ \overline{Fx}$ and all formulas $F(i),$
$i=1,,$

{\em def 8a:} $k$ is $\omega $-coherent iff: $Not(a)$ such that: $%
(n)[Sb(a_{Z(n)}^\nu )\in Flg(k)]\&[Not(\nu Gena)]\in Flg(k)$

that can be read in these terms: it doesn't exist any sequence, or sign of
class $a$ for which the substitution in $a$ of the numeral of $n$ to the
free variable of a can be interpreted as a consequence of the $k$%
-propositions and is not a generalization of $\nu $ respect to $a$ .

{\bf Theorem G-VI } - following G\"odel \cite{godel}

for each class $k$ of $\omega $-coherent recursive formulas exist signs of
recursive classes $r$ such that $(\nu Genr)\&\symbol{126}(\nu Genr)\in
Flg(k) $, i.e. undecidable propositions do exist inside a $\omega $-coherent
structure.

{\bf Proposition p1:} each formal system $S$ (as Dirac adimensional
construction) containing $Z$ with a finite number of axioms and having as
inference principles the rule of introduction and that of implication is not

complete: there do exist undecidable propositions starting from $S$ axioms,
if $S$ is non-contradictory by the following

{\bf Proposition p2:} in $S$ is not possible to show that the proposition
asserting the non-contradiction of $S$ itself is true or false.

Those theorems hold also for systems with an infinite number of axioms and
different inference principles, unless all formulas are sorted and numbered
e.g. by length (etc.) as in a computer database, also the classes of numbers
associated to axioms. Is also possible to expand the system by introducing
some variables for number classes, for classes of number classes, and so on,
with their understanding axioms, till going transfinite formal systems where
theorems hold, and $\omega $-non contradictoriety in one of those systems is
demonstrable inside other bigger systems. Also non decidable propositions
that demonstrate {\bf p1} became decidable if are introduced into higher
level structures with their axioms, that are inevitably affected by other
undecidable propositions, and so on, as is found in theory of sets. Building
all G\"odel construction for the universe, the measure is easily shown to be
a $k$-class of $\omega $-coherent propositions in terms of event
meta-classes, that contains undecidable propositions inside the structure
universe, in fact all logical structures can be took back to the above basic
formal construction. This is assured by:

{\bf Theorem} {\bf 1a:} (see G\"odel \cite{goe1}) there is no realization
with a finite number of elements for which all and only demonstrable
formulas in a system of logical propositions H are satisfied, giving
privileged values for each substitution.

{\bf 1b:} between H and the system A of usual propositional calculus there
is an infinite number of systems, i.e. $\exists $ a decreasing monotonic
sequence of system such that each of them contains H and is contained in A.

\ {\bf Lemma:} the measurement process, in the adimensional construction, is
a $k$-class of $\omega $-coherent propositions and contains undecidable
propositions.\ 

{\bf proof:} {\it reductio ad absurdum:} supposing exists $a$ such that 
$(n)[Sb(a_{Z(n)}^\nu )\in Flg(k)]\&[Not(\nu Gena)]\in Flg(k)$, it means that
substituting for each event $n$ its numeral $Z(n)$ to the free variable $\nu 
$ of the sequence of events intended as formulas and measurements, it
becames a direct consequence of those $k$ rules (assertions or formulas)
without being a generalization of $a$ free variables, i.e. without being an
extension of the system measure-events. But, during evolution, by state
coherence seen above, must change the structure of the system
observer-enviroment observed \cite{zur,zur1,zur2}, that written in the symbolic
language corresponds to: $Not(\nu Gena)\notin
Flg(k)\Rightarrow Not(\exists )a$.

In fact, a measure gives relationships between other events created by the process itself, i.e. the sequence $a$ representing the measurement is a causal sequence of events or coincidence of events that can also belong to the class of ``laws'': the evolution of state universe is a change of the structure universe itself and must be a generalization of the above structure as a self-reproducing fractal one. 
For this reason, Everett's Many-Worlds construction is a metastructure for the measurement process and the
problem of measurement is translated into the problem of a choice inside a Many-Worlds scenario. The metamathematical structure representing those universes, as it happens for every type of formal logic system based on integer numbers, can be translated into a meta structure admitting a splitting that is to be interpreted as true or false. This means that the $M$ Rovelli operator \cite{rovelli} is defined and gives a projection of the state ``{\it happened}'' or not happened for the universe from all its potential possibilities of existence. Proposition measure and time evolution are decidable only in this metastructure, and the measure ''Universe'' is seen as a k-proposition justifiable only inside this meta-universe, defined by mutual relationships between subsets of the universe itself. In terms of G\"odel numbers it is represented by the smallest one that defines the next universe chosen between other universe states: in fact by definition of k-structure, it is the smallest set of information that has all and only these informations needed to generate that $f-$state of the universe that we name ``{\it future}''. 
All this means that if the structure Universe so defined should be logically coherent, contains some undecidable propositions that can be neither true nor false, suggesting the use of self bootstrapped models.

It is so possible to build, following those prescriptions, a metastructure in terms of G\"odel numerals, and by Church's lemma, a code for Quantum Turing Machines in the usual Heisenberg representation for quantum computers able to describe the universe inside Many-Worlds Everett scenario \cite{proof13}. 
Time evolution can be set as the problem for self-reproducing a semantically closed cellular automata, that evolves defining space, time, matter and energy as its own aspects, and draws its existence by mutual interaction of its parts, as a self-bootstrapped net does, modifying its own structure during evolution. 
This implies a self-referential linguistic mechanism, which description is based on symbols related to physical structures or internal states, as von Neumann automaton \cite{neum} does: a self-replicating scheme with memory stored description $\Phi (A)$, which can be interpreted by the universal constructor $A$ to produce $A$ itself; in addition there is an automaton $B$ capable of copying any description $\Phi$ included into the self replication scheme and a third automaton $C$ for manipulation of description, $\circ$. According to Quantum Mechanics prescriptions the initial state $A$ must be destroyed in a mixed one with an ancilla $C$ to be reproduced in a new state.

The self replicating system is structured into the set of automata $(A+B+C)$ representing the metamathematical description of the universe, and the semantical description $\Phi (A+B+C) \rightarrow (A+B+C) \circ $ is needed to construct the new automaton and describe the new - possible - state of the Universe. A system like this, which is able to relate internally stable structures to an interaction with an environmental metastructure, could be seen as a self-reproducing ''{\it organism}'' with its semantic closure, and the code mathematically maps ''{\it instructions}'', (which are physical actions) into physical actions, to be performed by its composition rules including emergent physical and linguistic structures.

An alternative description could be given by the self-referential systems that, instead, have different logical constructions that are used for the description of the Universe. This was shown with an end-game modeling \cite{cah,cah2,kitto,kitto2}, presenting many properties as a fractal 3- space, derived by universality property; these structures make possible fundamental interactions modeling as a fractal-like structure of emerging spacetime.
The description of finite time Quantum Turing Machines corresponds to the use of Heraclitean processes for self-organizing critically systems, start up axioms are suppressed by requiring that the logic must be self-consistently bootstrapped.
In this vision, the system moves itself into states characterized by a stochastic process that follow a fractal-like description with no fundamental scales expressed with the language of Wiener processes and shortening all fundamental processes in a very compact way \cite{proof13}.

\section{Conclusions}

In our formal language the Universe can be described in different ways. Each single bit of information is the result of a physical process inside the Universe.
An example is an end game modeling, a self-referential system with a semantically closed structure containing intrinsic randomness as any of its representations is a subset where there is not a global object-environment loss of decoherence but interactions with its subsets indicating that its evolution including the mathematical truths, undecidable propositions and quantum measurement problem are metastructures inside the Universe built with interactions between subsets.
Our formal language is based on mathematical truths and any formal mathematical modeling of the Universe as a whole as initially proposed by Dirac inherits G\"odel's undecidability. 
If one adopts this approach, the evolution and the problem of measurement is deeply related to undecidable propositions and Everett's Many-Worlds interpretation of Quantum Mechanics becomes a meta-structure to contain all the possible states of the past and future evolution, showing the limits of our formal language in the description of the Universe as a whole, showing a clear difference between models of physical systems and a complete formal language.
Following Ref. \cite{tipler2021}, the language of Many-Worlds Quantum Mechanics is different from that of Quantum Mechanics where any event depends on the probability amplitude of the wavefunction of any possible event. In our case, if we have to adopt the Many-Worlds scenario, in this case ``the wave function is a relative density of universes in the multiverse amplitude''. This means that in Many-Worlds, the Born frequencies, related to the square of the absolute value of the wavefunction that gives the best estimate of the probability density is caused by the intrinsic deterministic nature of the wave equation. This implies, as described by our deterministic classical language that the evolution of quantum systems in the multiverse, the Universe of universes, is described by a deterministic wave equation. Born frequencies are thus approached, asymptotically, rather than being defined a priori as occurs in the computation of the Quantum Turing Machine discussed before, which can be isomorphically related to a classical Turing Machine. From this, also the nonlocality of Quantum Mechanics has a different interpretation from that in the Many-Worlds \cite{tipler2}, suggesting that at the Planck's scale there will be not a different behavior, any observer should be ``quantum'' or a classical singularity. Instead, from the properties discussed at those scales \cite{tambu1} and near the horizon of a black hole \cite{tambu2}, this would lead unavoidably to a holographic scenario described by cellular automata \cite{tambu3}, which founds an agreement with the Many-Worlds scenario when handled with our language and mathematical truths. 
Being the description of the Universe built with mathematical truths, and identifying the set of possible Universes of Everett's scenario as a class containing each possible set of events-subsets, this defines the Universe like in the Russell's paradox.
The only way is to postulate the independent existence of the so-called laws of physics and consider any pure mathematical modeling for the description of the Universe incomplete.
The limits of our language suggest that we cannot use it to simulate a Universe as it requires truths from axioms acting in terms of ``outside'' and followed by inference laws with theorems and undecidable propositions. The only way one could simulate the Universe is to use a semantically closed structure based on a quantum language that is not accessible for an observer (like us) as it acts with its complementary in the Universe needing a free will axiom in Quantum Mechanics \cite{tofft,tofft2}.



\begin{thebibliography}{999}
\bibitem{proof11}  {\bf proof:} given the formula $F_n:\sum (a_i\supset
\subset a_k)$ with $1\leq i<k\leq n$ and $\sum $ be the iterated disjunction 
$\vee $ and $a_j$ are propositional variables. $F_n$ is satisfied by every
realization with less than $n$ elements, i.e. are satisfied all demonstrable
formulas in H. In fact, for each substitution, in at least one of the
components for $F_n,a_i,a_k$ will be substituted with the same element $e$;
and $e$ $\supset \subset e.\vee b$ for arbitrary $b$ gives a privileged
value, because $e$ $\supset \subset e.\vee b$ is H-demonstrable. Taking e.g.
the realization $S_n$

elements: $\{1,2,...n\},$ privileged element $1$;

$a\vee b=\min (a,b);\ a\wedge b=\max (a,b);\ a\supset b=1$ if $a\geq b;$

$a\supset b=b$ for $a<b;\ \neg a=n$ for $a\neq n,\ \neg n=1.$

In this case are satisfied by $S_n$ all H formulas and $F_{n+1}$ formula
with all $F_j;\ j>n$; but all formulas with smaller $j<n$ values are not
satisfied$\Rightarrow $no $F_n$ formula is demonstrable in H.
\bibitem{proof13}  This is assured by a direct consequence from G\"odel
conjectures stating that for every formal structure it does exist a
metastructure where some undecidable propositions became decidable
(logically true or false) and other are described in a more compact way, as
the fractal structure assures, employing as main fractal generator the self
similarity on the structure given by Heisenberg uncertainty principle. Is
also possible to obtain a G\"odel code for lie algebras, quantum computer
with self referential noise, superspace tools, etc by Church's lemma: in
some works \cite{winter,chuang} is shown that first order one error quantum computers can give Lie, C*-algebra structures, and a good representation of physical modeling. This corresponds to a finite time computation in a classical
Turing Machine given by the finiteness of the quantum of action in
Heisemberg uncertainty principle, related to Quantum Turing Machine halting problem \cite{linden,gottesman}.
\bibitem{proof5b}  
perpetuum mobile: considering statistical operators for
composed systems before and after a measurement, $\rho _b^{comp}=P_{\psi
_b}^{comp};$ $\rho _a^{comp}=\sum \omega _iP_{\psi _{a_i}}^{comp},$ the mean value of the energy is expressed in terms of the Hamiltonian H: $E=Tr\{\rho^{comp}H\}$ and the energy and entropy change are $\Delta E=\sum_i\omega_i(\Psi _{ai},H\Psi _{ai})-(\Psi _b,H\Psi _b);\ \Delta \sigma =-\sum_i\omega_i\ln \omega _i>0$. If an observable $A$ is measured, this means that $\Delta E\neq 0$ violating energy conditions. But by thermodynamics laws $
\Delta E=Q+W=tr\{(\Delta \rho ^{comp})H\}+Tr\{\rho ^{comp}(\Delta H)\}$ that
gives for a qjump $\Delta E=Q;$ but $Q>0,\ \Delta \sigma >0$ that is a
self-heating system.
\end{thebibliography}
\end{document}